\documentclass[slac_one]{revtex4}
\usepackage{graphicx}
\usepackage{fancyhdr}
\begin{document}

\title{{\small{2005 International Linear Collider Workshop - Stanford,
U.S.A.}}\\ %% Please keep this conference title here
\vspace{12pt}
Spatial Resolution of a Micromegas-TPC Using the Charge Dispersion Signal} %% Paper title goes here


\author{A.~Bellerive, K.~Boudjemline, R.~Carnegie, M.~Dixit, J.~Miyamoto, 
% H. Mes, 
E.~Neuheimer, A.~Rankin, E.~Rollin, K.~Sachs} 
\affiliation{Carleton University, Ottawa, Canada}

\author{J.-P. Martin}
\affiliation{University of Montreal, Montreal, Canada}

\author{V. Lepeltier}
\affiliation{LAL, Orsay, France}

\author{P. Colas, A. Giganon, I. Giomataris}
\affiliation{CEA, Saclay, France}

\begin{abstract}
\setlength{\unitlength}{1mm}
\begin{picture}(0,0)
\put(120,88){\parbox[t]{5cm}{\normalsize
Carleton Phys0508\\
DAPNIA 05-232 \\ LAL 05-99}}
\end{picture}%
The Time Projection Chamber (TPC) for the International Linear Collider will
need to measure about 200 track points with a resolution close to 100 $\mu$m. A
Micro Pattern Gas Detector (MPGD) readout TPC could achieve the desired
resolution with existing techniques using sub-millimeter width pads at the
expense of a large increase in the detector cost and complexity. We have
recently applied a new MPGD readout concept of charge dispersion to a
prototype GEM-TPC
and demonstrated the feasibility of achieving good resolution with pads
similar in width to the ones used for the proportional wire TPC. The charge dispersion
 studies were repeated with a Micromegas
TPC amplification stage. We present here our first results on the Micromegas-TPC
resolution with charge dispersion. The TPC resolution with the Micromegas
readout is compared  to our earlier GEM results and to the resolution expected
from electron statistics and transverse diffusion in a  gaseous TPC. 
\end{abstract}

\maketitle

\thispagestyle{fancy}


\section{INTRODUCTION} % Section title should be in all capitals.
The Time Projection Chamber (TPC) for the future International Linear Collider
will need to measure about 200 track points with a resolution of better than 100
$ \mu$m. The resolution goal, close to the fundamental limit from ionization
electron statistics and transverse diffusion in the gas, is nearly two times better
than what has been achieved by conventional wire/pad TPCs.  A TPC with a Micro
Pattern Gas Detector (MPGD) readout could, in principle, reach the target
resolution. However, it may require sub-millimeter wide pads resulting in a
large increase in the number of electronics channels, detector cost and
complexity over conventional TPCs. 

We have recently developed a new concept of position sensing from charge
dispersion in MPGDs with a resistive anode \cite{Ref1}. With charge dispersion
wide readout pads similar in width to the ones used with proportional
wire/cathode pad TPCs can be used without sacrificing resolution. This was
demonstrated  recently with cosmic ray tracks for a TPC read out with a GEM 
with a resistive anode  \cite{Ref2}. We present here new results on the cosmic
ray track resolution of a TPC read out with a  Micromegas instrumented for charge 
dispersion with a resistive
anode. The GEM and the Micromegas TPC resolution measurements made with the
resistive readout are compared to our earlier GEM-TPC resolution results with a
conventional readout \cite{Ref3} and  to the expected resolution from transverse
diffusion and ionization electron statistics in a gaseous TPC.

\section{MEASUREMENT SETUP AND ANALYSIS}

\begin{figure*}[t]
\centering
\includegraphics[width=175mm]{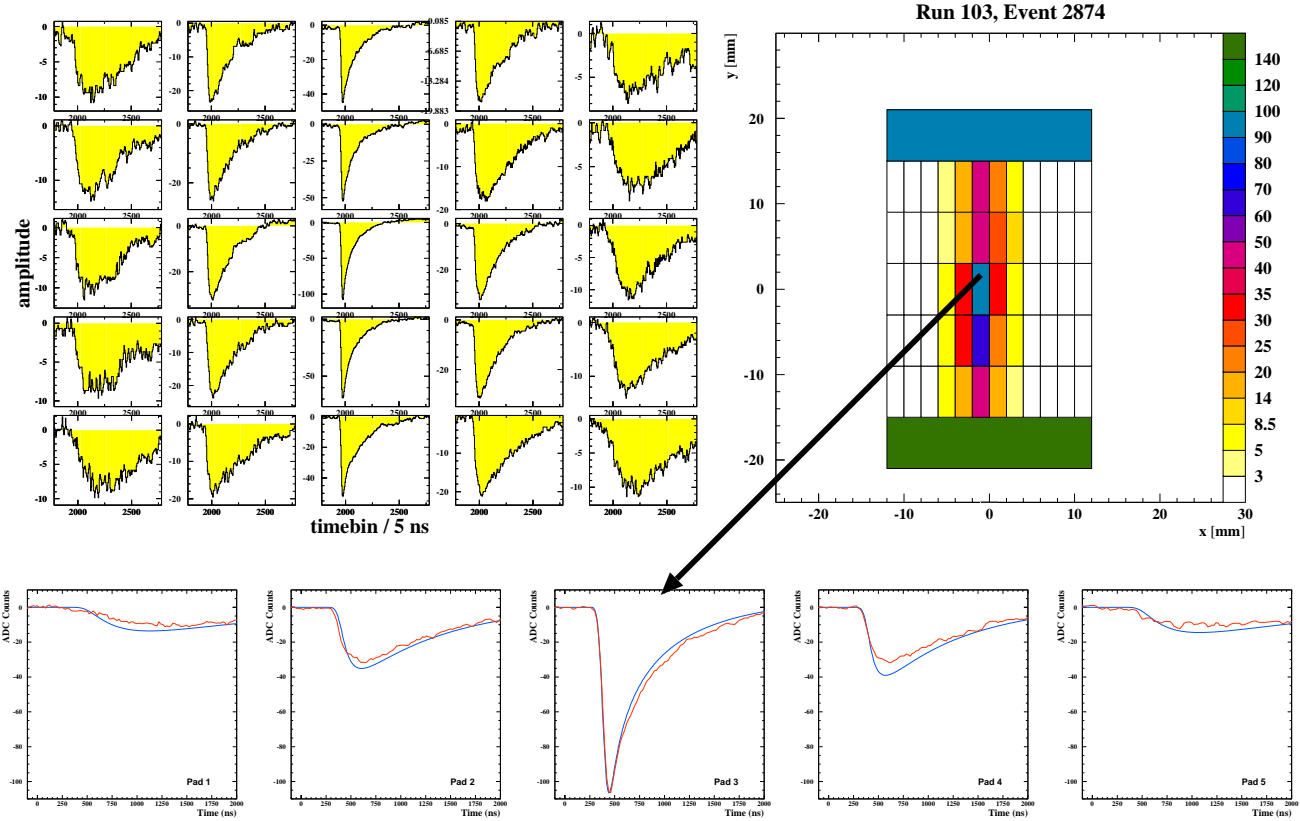}
\caption{Pad layout and observed signals for a cosmic ray track in a GEM-TPC
with a resistive anode readout. Also shown are simulated signals for the central
row of pads.  Detailed model  simulation includes  longitudinal and transverse
diffusion, gas gain, detector pulse formation, charge dispersion on a resistive anode and
preamplifier rise and fall time effects.} \label{Simulation} 
\end{figure*}

A small 15 cm drift length test TPC used earlier  with a conventional
direct charge  GEM readout  \cite{Ref3} was modified for charge dispersion
studies. The readout endcap was modified such that it could  accommodate either a
GEM or a Micromegas with a resistive anode \cite{Ref2} readout system.  The 
gas used, Ar:CO$_2$/90:10, was chosen to simulate the reduced transverse diffusion
conditions of a TPC in a magnetic field. Charge signals on 60 pads, 2 mm
x 6 mm each (see Figure \ref{Simulation}), were read out using Aleph wire TPC
preamplifiers and digitized directly  using 200 MHz custom built 8 bit FADCs. 

% If in two-column mode, this environment will change to single-column
% format so that long equations can be displayed. Use
%\begin{widetext}
%\end{widetext}

\begin{figure*}[t]
\centering
\includegraphics[width=85mm]{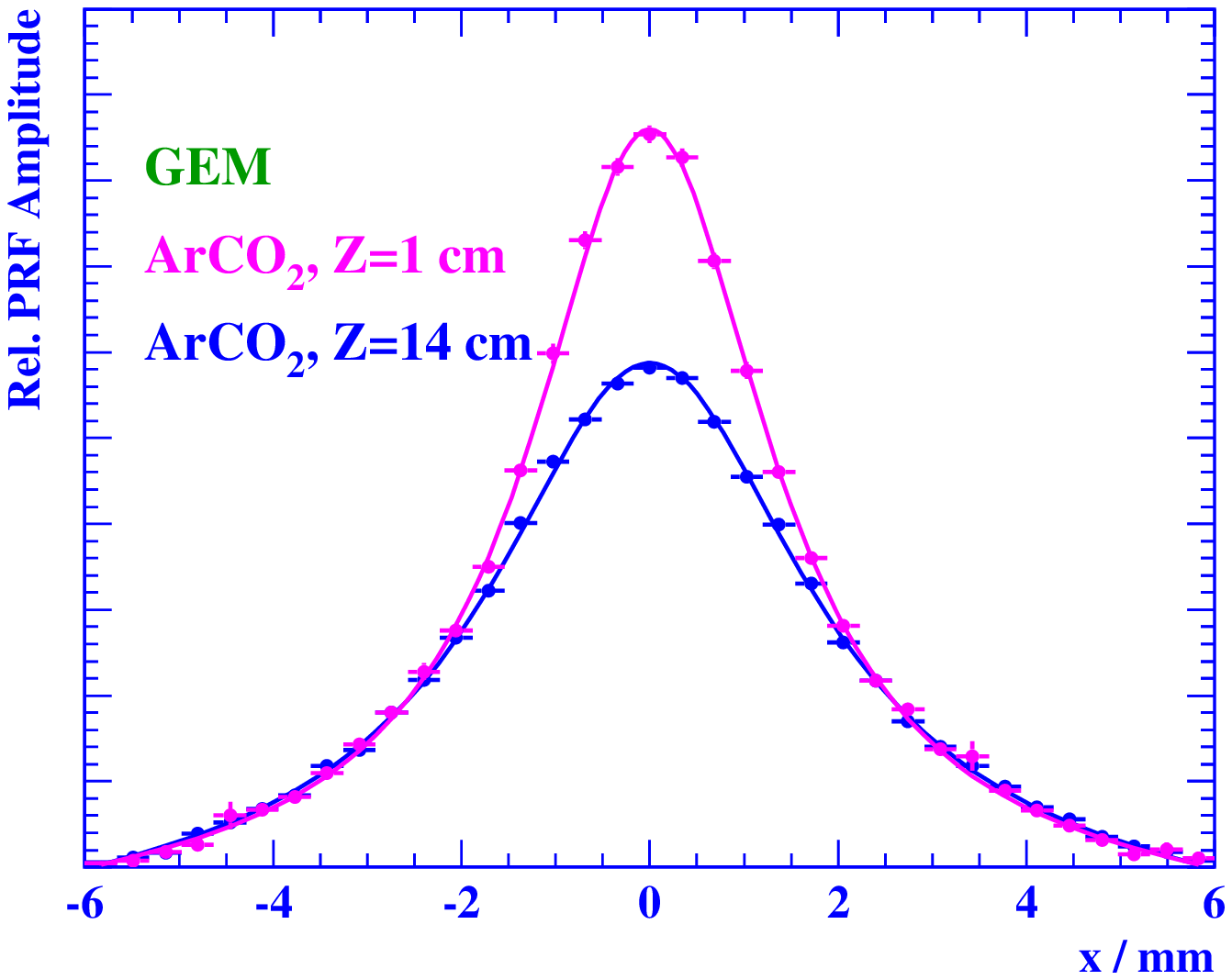}
\includegraphics[width=85mm]{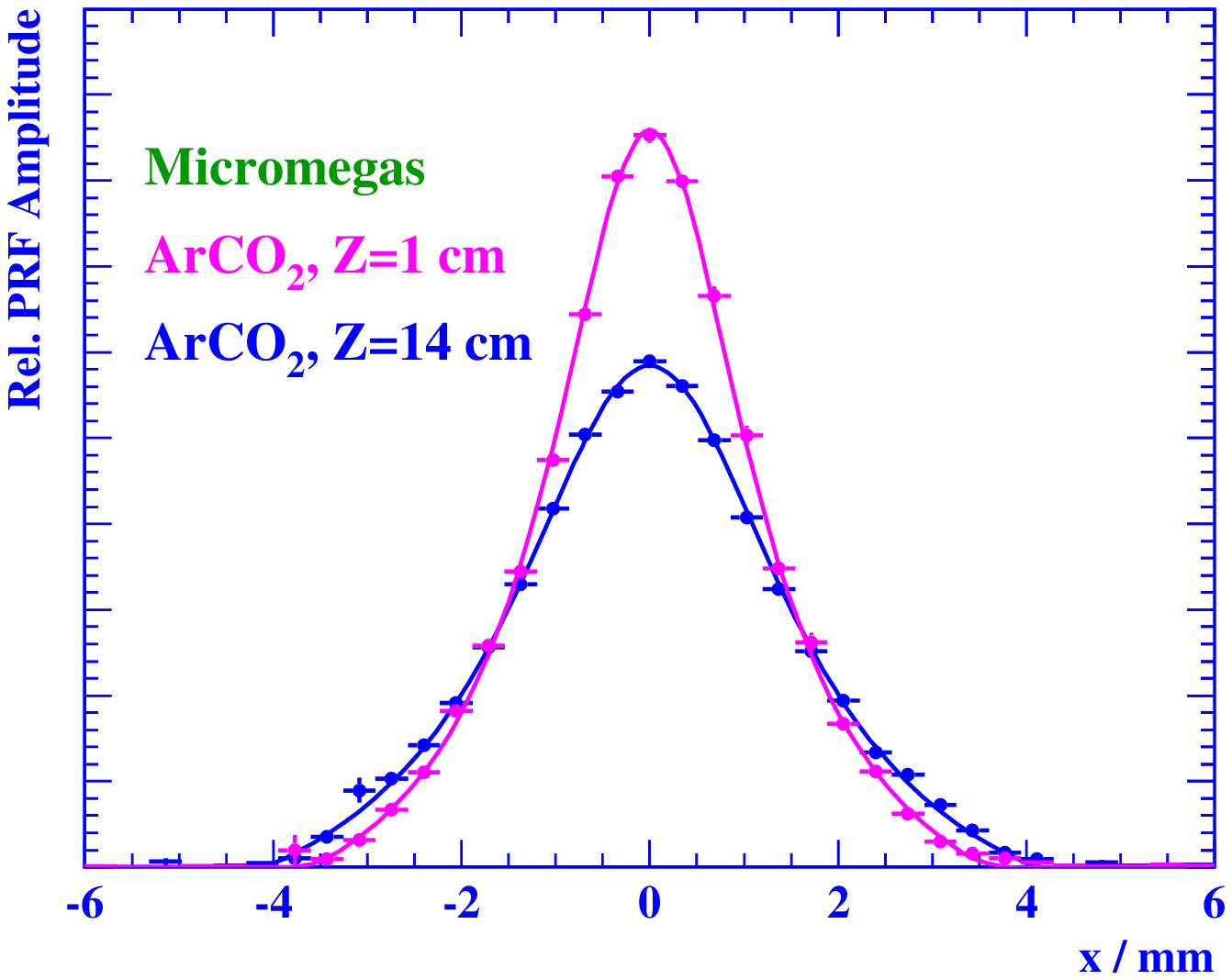}
\caption{Examples of the pad response function (PRF). The PRFs  were determined
from a subset of the cosmic ray data set. The PRF peak for
longer drift distances is lower due to Z dependent normalization. Compared to the GEM, the 
PRF width for the Micromegas
is narrower due to the use of a higher surface resistivity anode and smaller diffusion after gain.}
\label{PRFs}  
\end{figure*}

Track reconstruction  techniques used for the conventional direct charge
GEM-TPC readout \cite{Ref3} cannot be used   for the resistive anode MPGD
readout since non charge collecting pads nearby also have measurable signals. 
Not only the observed amplitude, but also the pulse shape depends on the relative 
location of the pad 
with respect to the track path. In theory, a first principle determination
of the track PRF is possible.    For the present, we have chosen to determine
the PRF empirically from the internal consistency of cosmic ray track data.   

For the purpose of analysis, the cosmic ray data were divided into two parts; 
one used for calibration and the other for resolution studies.  We use the
calibration data set to determine the PRF associated with the pads in a given
row. The PRFs were determined for 15 seperate 1 cm wide regions in drift
distance.  Figure \ref{PRFs} shows examples of PRFs for the GEM and the Micromegas
readout. 
The measured PRFs have been parameterized  with a ratio of two symmetric 4th
order polynomials: 
\begin{equation}
PRF(x,\Gamma,\Delta,a,b ) = \frac{1 + a_2 x^2 + a_4 x^4}{1 + b_2 x^2 + b_4 x^4} \; ,
\label{eq:prf}
\end{equation}
The  coefficients of the two 4th order polynomials $a_2$ and $a_4$, and  $b_2$ and $b_4$ can
be expressed in terms of the FWHM  $\Gamma$, the base width $\Delta$ of the
PRF, and two scale defining parameters a and b. 

\begin{figure*}[t]
\centering
\includegraphics[width=85mm]{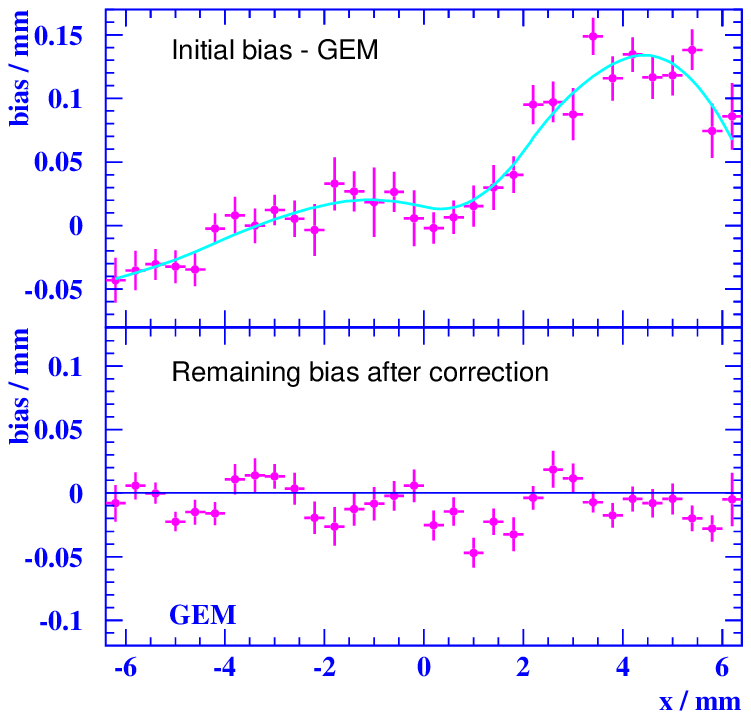}
\includegraphics[width=85mm]{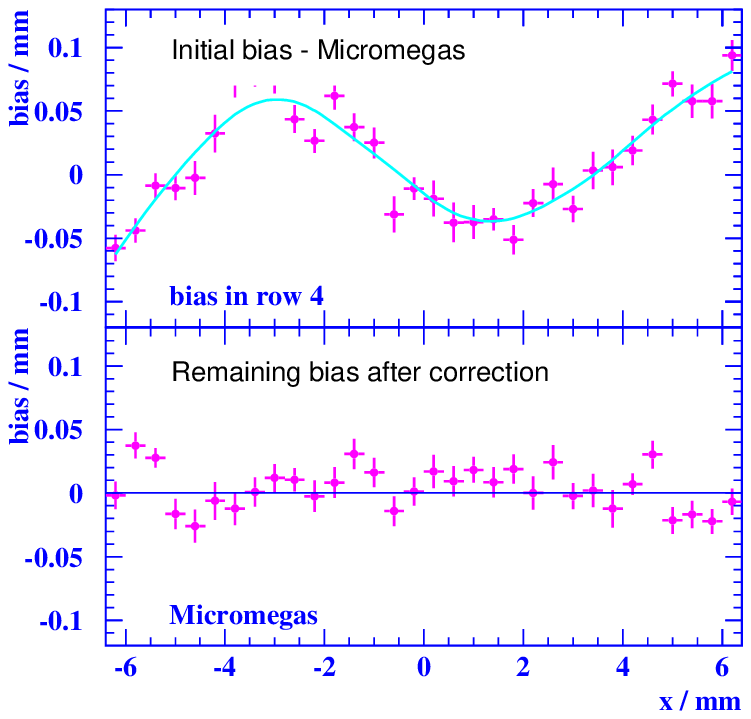}
\caption{Bias corrections for pads in row 4 for the GEM and the Micromegas resistive anode TPC cosmic ray data.
The lower set of figures show the remaining bias after correction.}\label{bias}  
\end{figure*}

The track fitting parameters $x_0$ and $\phi$ are
determined by fitting the PRF to the pad amplitudes for the full event by 
$\chi^2$ minimization. 
The position in a row
$x_{\rm row}$ is determined from a seperate one-parameter track fit 
to this
row only using the known track angle $\phi$. Figure \ref{bias} 
shows examples of bias in position determination with GEMs and Micromegas. The bias is 
the mean difference $x_{\rm row}-x_{\rm track}$ as a function of
$x_{\rm track} = x_0 + \tan{\phi}*y_{\rm row}$, where $y_{\rm row}$ is
the $y$ position of the row. A bias of up to 150 $\mu$m is observed
which may be attributed to a non-uniform RC constant due to inhomogeneities
in the gap size and the resistivity of the foil.
However, this bias is due to geometry only and can easily be corrected.
Figure \ref{bias} also shows the remaining bias after correction,
which is negligible.

\begin{figure*}[t]
\centering
\includegraphics[width=57mm]{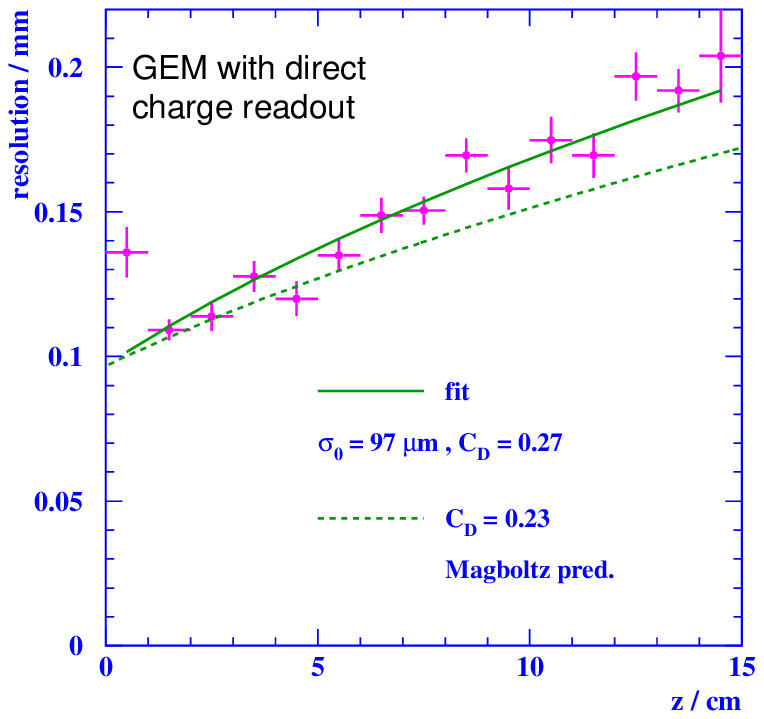}
\includegraphics[width=57mm]{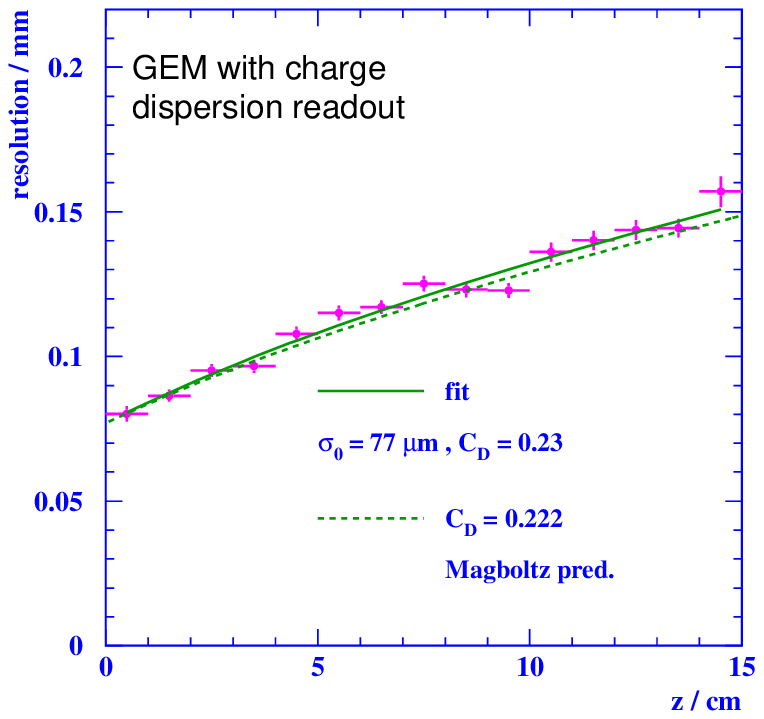}
\includegraphics[width=57mm]{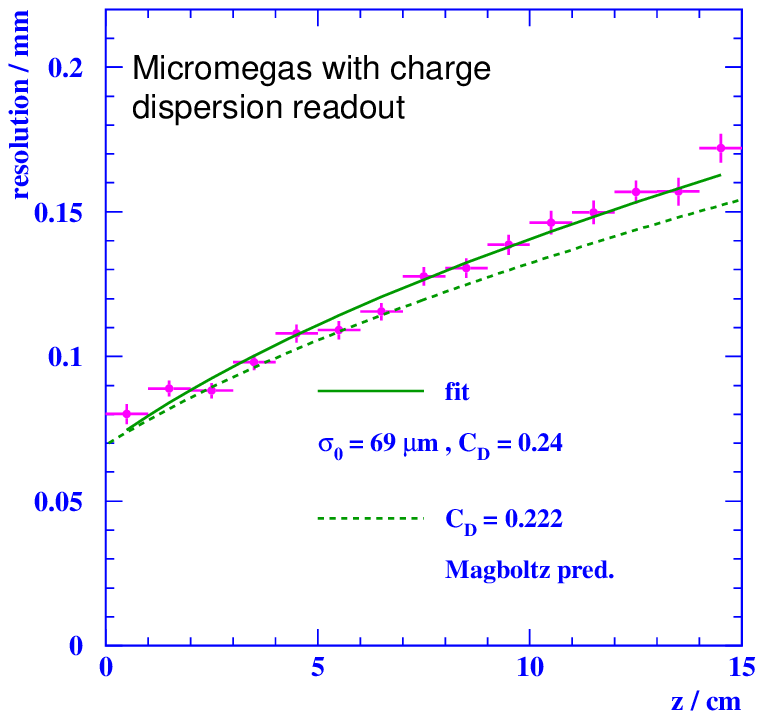}
\caption{Charge dispersion improves TPC resolution over that from direct charge
for low diffusion gases like Ar:CO$_2$ with limited charge sharing between pads.
Compared to direct charge readout, resistive readout gives better resolution for
the GEM and for the Micromegas both.}\label{Res1} 
\end{figure*}

Figure~\ref{Res1} shows the measured resolution  for Ar:CO$_2$/90:10. 
With GEMs we have results with and without the resistive anode; for Micromegas we 
have results with the resistive anode only. The resolution is fitted
to the function: 
\begin{equation}
s = \sqrt{ s_0^2 + C_{\rm D}^2/N_{\rm eff} \ast z} \; ,\label{eq:reso}
\end{equation}
where $s_0$ is the resolution at zero drift distance $z$. Here $C_D$ is the
diffusion constant and  
$N_{\rm eff}$ is the effective number of track ionization electrons over the length of a pad.  
$N_{\rm eff}$ is not the average number of electrons. Rather, it is given by:
$N_{\rm eff} ={1/ \langle \sqrt{1/N}  \rangle}^2$,
where N is the number of electrons following the Landau distribution. N is determined
from the measured pad-signal amplitudes scaled using the most probable number of track
ionization electrons obtained from simulation. 

Electronic noise and remaining systematic effects contribute to the constant
term $s_0$. The constant term $s_0$  is about 75$ \mu$m for the resistive
readout for both the GEM and the Micromegas. It can be compared to the larger
97$ \mu$m constant term for the normal GEM readout. 

As shown in Figure \ref{Res1},  the TPC resolution obtained with the resistive
anode readout for both the GEM and the Micromegas  is better than our previous
result with a conventional GEM readout \cite{Ref3}, where at long drift distance
the observed resolution  
was about 40\% worse than expected.
Apart from the constant term, the dependence of resolution on drift distance
with the resistive readout  
follows the expectation from transverse diffusion in the gas and electron statistics.

\section{SUMMARY AND OUTLOOK}

In summary, the charge dispersion on a resistive anode improves the MPGD-TPC
resolution significantly  over that achievable with conventional direct charge
readout for 2 mm wide pads. Bias errors due to  RC inhomogeneities can be
corrected. With no magnetic field, the measured dependence of resolution  on
drift distance follows the behavior expected from transverse diffusion in 
the gas.

\begin{acknowledgments}
We wish to thank Ron Settles for lending us the ALEPH wire TPC
amplifiers  used in these measurements.
The research was supported by a project grant from the 
Natural Science and Engineering Research Council of Canada and the Ontario Premier's
Research Excellence Award (IPREA).
\end{acknowledgments}

\end{document}